\documentclass[11pt]{article}
\usepackage[fleqn]{amsmath}
\usepackage{amssymb}
\newtheorem{theorem}{Theorem}[section]

\begin{document}


\title{A Few More Quadratic APN Functions}

\author{Carl Bracken \thanks{Research supported by
Irish Research Council for Science, Engineering and Technology
Postdoctoral Fellowship} 
\&
Eimear Byrne\thanks{Research supported by the Claude
Shannon Institute, Science Foundation Ireland Grant 06/MI/006} \& Nadya Markin\thanks{Research supported by the Claude
Shannon Institute, Science Foundation Ireland Grant 06/MI/006} \& Gary McGuire\thanks{Research supported by the Claude
Shannon Institute, Science Foundation Ireland Grant 06/MI/006}\\  \\
School of Mathematical Sciences\\   University College Dublin\\   Ireland\\}


\maketitle

\begin{abstract}
\noindent We present two infinite families of APN functions on
$GF(2^n)$ where $n$ is divisible by $3$ but not $9$.
Our families contain two 
already known families as special cases.
We also discuss the inequivalence proof (by computation) 
which shows that these functions are new.
\end{abstract}


\bigskip



\bigskip


\section{Introduction}

Let $L=GF(2^n)$ for some positive integer $n$.
A function $f : L \longrightarrow L$ is said to be \emph{almost perfect nonlinear} (APN) on $L$
if the number of solutions in $L$ of the equation
$$f(x+q)+f(x)=p$$
is at most 2, for all $p,q\in L$, $q \not=0$. Equivalently, $f$ is APN if the set
$\{f(x+q)+f(x): x \in L \}$ has size $2^{n-1}$ for each $q \in L^*$. 
Clearly, as $L$ has characteristic 2, the number of solutions
to the above equation must be an even number for any function $f$ on $L$.

APN functions were introduced in \cite{N} by Nyberg, who defined them as the mappings with
highest resistance to differential cryptanalysis. In other words, APN functions are those for which
the plaintext difference $x-y$ yields the ciphertext difference $f(x)-f(y)$ with probability $1/2^{n-1}$.
Since Nyberg's characterization, many papers have been written on APN functions, although not many
different families of such functions are known.

\bigskip

The main result of this paper is a construction of a new family of APN functions.

\bigskip

Two functions $f,g : L \longrightarrow L$
are called {\em extended affine} (EA) equivalent
if there exist affine permutations $A_1,A_2$ and an affine map $A$ such that
$g=A_1\circ f \circ A_2 + A$.

Until recently, all known APN functions were
EA equivalent to one of a short list of monomial functions,
namely the Gold, Kasami-Welch, inverse, Welch, Niho and Dobbertin functions.
For some time it was conjectured that this list was the complete
list of APN functions up to EA equivalence.

A more general notion of equivalence has been suggested in \cite{CCZ},
which is referred to as Carlet-Charpin-Zinoviev (CCZ) equivalence.
Two functions are called CCZ equivalent if the graph
of one can be obtained from the graph of the other by an affine permutation
of the product space. EA equivalence is a special case of CCZ equivalence.

We say that $f:L \longrightarrow L$ is differentially $m-$uniform if the polynomial
$f(x+q)+f(x)+p$ has at most $m$ roots in $L$, for any $p,q\in L$, $q\not=0$. 
Then $f$ is APN on $L$ if and only if it is 
differentially 2-uniform on $L$. 
Differential uniformity, and resistance to linear and differential attacks, are 
invariants of CCZ equivalence.

In \cite{BCP}, Proposition 3, the authors express necessary and sufficient
conditions for EA equivalence of functions in terms of CCZ equivalence
and use this to construct several examples of APN functions that are
CCZ equivalent to the Gold functions, but not EA equivalent to any monomial function.
This showed that the original conjecture is false.
The new question was whether all APN functions
are CCZ equivalent to one on the list.

In 2006 a sporadic example of a binomial APN function
that is not CCZ equivalent to any power mapping was given in \cite{EKP}.
A family of APN binomials on fields $\mathbb{F}_{2^n}$, where $n$ is divisible
by $3$ but not $9$, was presented in \cite{BCFL}.
In \cite{BCL} these have been shown to be
EA inequivalent to any monomial function, and CCZ inequivalent to the Gold
or Kasami-Welch functions.
For the case $n=6$, in \cite{Dillon} Dillon presented a list of CCZ inequivalent APN functions on $GF(2^n)$, found by computer search.

Below we list all the infinite families of non-monomial APN functions known at the time of writing.
These families are all pairwise CCZ inequivalent.

\bigskip

\begin{enumerate}
\item
$$f(x)= x^{2^s+1}+ \alpha x^{2^{ik}+2^{mk+s}},$$
where $n =3k$, $(k,3)=(s,3k)=1$, $k \geq 3$, $i \equiv sk \mod 3$, $m \equiv -i \mod 3$, $\alpha = t^{2^k-1}$ and $t$ is primitive (see 
 Budaghyan,  Carlet,   Felke,  Leander \cite{BCFL}).

\bigskip

\item

$$f(x)= x^{2^s+1}+ \alpha x^{2^{ik}+2^{mk+s}},$$ where 
$n =4k$, $(k,2)=(s,2k)=1$, $k \geq 3$, $i \equiv sk \mod 4$, $m= 4-i$, $\alpha = t^{2^k-1}$ and $t$ is primitive (see Budaghyan,  Carlet,  Leander \cite{BCL2}).
This family generalizes an example found for $n=12$ by
 Edel, Kyureghyan, Pott \cite{EKP}.

\bigskip

\item
$$f(x) = \alpha x^{2^{s}+1}+{\alpha}^{2^k}x^{2^{k+s}+2^k}+\beta
x^{2^k+1}+\sum_{i=1}^{k-1} {\gamma}_i x^{2^{k+i}+2^i},$$
where $n=2k$,
$\alpha$ and $\beta$ are primitive elements of $GF(2^{n})$, and
${\gamma}_i \in GF(2^k)$ for each $i$, and $(k,s)=1$, $k$ is odd,
$s$ is odd (see  Bracken,  Byrne,  Markin,  McGuire \cite{BBMMcG}).

\bigskip

\item
$$f(x) = x^3 + Tr(x^9),$$
over $GF(2^n),$ any $n$ (see Budaghyan,  Carlet,  Leander  \cite{B+}).

\bigskip

\item
$$f(x)= ux^{2^{-k}+2^{k+s}}+u^{2^{k}}x^{2^{s}+1}+vx^{2^{k+s}+2^{s}},$$
where $n=3k$,
$u$ is primitive, $v \in GF(2^k), (s,3k)=1,\  (3,k)=1$ and 3 divides $k+s$
(see Bracken,  Byrne,  Markin,  McGuire \cite{BBMMcG}). 

\bigskip

\item
$$F(x)=u^{2^k}x^{2^{-k}+2^{k+s}}+ux^{2^{s}+1}+vx^{2^{-k}+1}$$
where $n=3k$,
$s$ and $k$ are positive integers with $k+s$ divisible by three and
$(s,3k)=(3,k)=1$,  $u$ is a primitive element of ${GF(2^{3k})}$ and  $v \in
GF(2^{k})$  (this paper).

\bigskip

\item
$$F(x)=u^{2^k}x^{2^{-k}+2^{k+s}}+ux^{2^{s}+1}+vx^{2^{-k}+1}+wu^{{2^k}+1}x^{2^{k+s}+2^s}$$
where $n=3k$,
$s$ and $k$ are positive integers with $k+s$ divisible by three and
$(s,3k)=(3,k)=1$,  $u$ is a primitive element of ${GF(2^{3k})}$ and  $v,w \in
GF(2^{k})$ with $vw \neq 1$ (this paper).

\end{enumerate}

In general, establishing CCZ equivalence of arbitrary functions is extremely difficult. There are, however, a number of invariants of CCZ equivalence that can be useful in the classification of functions. A nice link with coding theory is that a pair of functions $f$ and $g$ on $L$ are CCZ equivalent on $L$ if and only if the binary codes with parity check matrices 
$$H_f = \left[ \begin{array}{cccc} 
1&\cdots&1\\
 {\bf x}_1 & \cdots &{\bf x}_{2^n}\\  
f({\bf x}_1) & \cdots & f({\bf x}_{2^n})\end{array} \right] ,\:\: 
H_g= \left[ \begin{array}{ccc}  
1&\cdots &1\\
{\bf x}_1 & \cdots &{\bf x}_{2^n}\\   g({\bf x}) & \cdots & g({\bf x}_{2^n})\end{array} \right] $$
are equivalent over $GF(2)$,
see \cite{BBMMcG}.
Here ${\bf x}_i,f({\bf x}_i)$ and $g({\bf x}_i)$ are expressions of $x_i,f(x_i)$ and $g(x_i)$ respectively as binary vectors of length $n$ in $L$ viewed as a $GF(2)$ vector space and 
$L = \{x_1,...,x_{2^n}\}$. 

\bigskip

In this paper we introduce a new family of APN functions
on fields of order $2^{3k}$ where $k$ is not divisible by 3. The family of polynomials has the form
\begin{equation*}
F(x)=u^{2^k}x^{2^{-k}+2^{k+s}}+ux^{2^{s}+1}+vx^{2^{-k}+1}+wu^{{2^k}+1}x^{2^{k+s}+2^s}
\end{equation*}
with certain constraints on the integers 
$s,k$ and on $u,v,w \in GF(2^{3k})$ (see Theorem \ref{thmain}, or Family 7 in the introduction).
Curiously, setting $w=0$ gives a different family of trinomial APN functions (Family 6,
see Section 3).

The layout of this paper is as follows.
In the next section we show that our polynomials  are indeed APN functions on $GF(2^{3k})$.
Using code equivalence, in Section 3 we explain the fact that these functions are not CCZ equivalent
to any known APN functions when $n=12$, and are therefore new.

\section{New APN functions}

The following theorem construct quadratic
quadrinomial APN functions on $GF(2^n)$
 whenever $n$ is divisible by 3 but not 9. A quadratic monomial is one of the form
 $x^{2^i+2^j}$  for some integers $i$ and $j$. Observe that if $f(x)=x^{2^i+2^j}$, then
 $$f(x+q)+f(x)+f(q) = x^{2^i}q^{2^j}+x^{2^j}q^{2^i}$$
 is a linear function in $x$, whose kernel has the same size as any of its translates, such as the solution set of $f(x)+f(x+q)=p$ in $L$, for any $p \in L$. 
Because of this property, proving whether or not a quadratic polynomial is APN is more tangible than one that is not quadratic. For this reason, all of the recently discovered families of APN functions have been quadratic.

 We will show that our polynomial $F(x)$ is APN by computing the size of the kernel of the corresponding linear map $$F(x+q)+F(x)+F(q).$$

\begin{theorem}\label{thmain}
Let $s$ and $k$ be positive integers with $k+s$ divisible by three and
$(s,3k)=(3,k)=1$. Let $u$ be a primitive element of ${GF(2^{3k})}$ and let $v,w \in
GF(2^{k})$ with $vw \neq 1$. Then the function 
$$F(x)=u^{2^k}x^{2^{-k}+2^{k+s}}+ux^{2^{s}+1}+vx^{2^{-k}+1}+wu^{{2^k}+1}x^{2^{k+s}+2^s}$$
is APN over $GF(2^{3k})$.
\end{theorem}

Proof:

We show that for every $p$ and $q$ (with $q \neq 0$) in $GF(2^{3k})$ the equation
$$F(x)+F(x+q)=p$$
has at most two solutions by counting the number of solutions to the equation
$$F(x)+F(x+q)+F(q)=0.$$
This gives
\begin{eqnarray*}
F(x)+F(x+q)+F(q) & = & u^{2^k}(x^{2^{k+s}}q^{2^{-k}}+q^{2^{k+s}}x^{2^{-k}})+u(x^{2^{s}}q+q^{2^{s}}x)\\
                                & + & v(x^{2^{-k}}q+q^{2^{-k}}x)+wu^{2^k+1}(x^{2^s}q^{2^{k+s}}+q^{2^s}x^{2^{k+s}})=0.
\end{eqnarray*}
Replace $x$ with $xq$ to obtain
\begin{eqnarray*}
u^{2^k}q^{2^{-k}+2^{k+s}}(x^{2^{k+s}}+x^{2^{-k}})+uq^{2^{s}+1}(x^{2^{s}}+x)
+  vq^{2^{-k}+1}(x^{2^{-k}}+x) \\ + wu^{2^k+1}q^{2^{k+s}+{2^s}}(x^{2^s}+x^{2^{k+s}})=0,
\end{eqnarray*}
and collect terms in $x$ to get
\begin{eqnarray*}
\Delta(x)  :=  (vq^{2^{-k}+1} +uq^{2^s+1})x+(vq^{2^{-k}+1}+u^{2^k}q^{2^{-k}+2^{k+s}})x^{-k} \:\:\:\:\:\:\:\\ 
                 +   (wu^{2^{k}+1}q^{2^{k+s}+2^s}  + uq^{2^s+1})x^{2^s} +(wu^{2^k+1}q^{2^{k+s}+2^s}+u^{2^k}q^{2^{-k}+2^{k+s}})x^{k+s}=0.
\end{eqnarray*}
We write 
$$\Delta(x)=Ax+Bx^{2^{-k}}+Cx^{2^s}+Dx^{2^{k+s}}$$ 
where 
\begin{eqnarray*}
A &= & vq^{2^{-k}+1}+uq^{2^s+1},\:\:\:\: 
B=vq^{2^{-k}+1}+u^{2^k}q^{2^{-k}+2^{k+s}},  \\
C & = & wu^{2^{k}+1}q^{2^{k+s}+2^s}+uq^{2^s+1},\:\:\:\:
D=wu^{2^k+1}q^{2^{k+s}+2^s}+u^{2^k}q^{2^{-k}+2^{k+s}}.
\end{eqnarray*} 
Clearly $0$ is a root of $\Delta(x)$. Moreover $\Delta(1) = A+B+C+D=0$. 
If we show that 0 and 1 are the \emph{only} solutions of $\Delta(x)=0$,  then
we will have proved that $F(x)$ is APN on $GF(2^{3k})$. 

First we demonstrate that none of $A, B, C $ or $D$
vanish for any $q \in GF(2^{3k})^*.$ If $A=0$ we have
$u=vq^{2^{-k}-2^s}$ which implies $u^{2^k}=vq^{1-2^{k+s}}$. 
By hypothesis, $k+s$ is divisible by 3, so that $1-2^{k+s}$ is divisible by 7,
 and hence $q^{1-2^{k+s}}$ is a $7$th power
in $GF(2^{3k})$. Since $3$ does not divide $k$, $7$ does not divide $2^k-1$, 
so the map $x \mapsto x^7$ is a permutation on $GF(2^k)$. Then $v \in GF(2^k)$
can be expressed as a 7th power. This means that $u^{2^k}$ and hence $u$ is a 7th power in 
$GF(2^{3k})$. This gives a contradiction as
7 is a divisor of $2^{3k}-1$ and we chose $u$ to be primitive in $GF(2^{3k})$. We deduce that 
$A \neq 0$. 
Similar arguments show that $B, C$ and $D$ are all nonzero. 

Next we define the linearized polynomial: 
$$L_\theta(T):=T+\theta T^{2^{k}}+{\theta}^{2^{k}+1}T^{2^{-k}}.$$
When $T=\theta x+ x^{2^{-k}}$ and $\theta$ is a $({2^{k}-1})$-th power,
a routine calculation verifies that $L_\theta(T)=0$ for all $x \in
GF({2^{3k}}).$ 
Observe that 
$$\frac{A}{B}=
\frac{vq^{2^{-k}+1}+uq^{2^s+1}}{vq^{2^{-k}+1}+u^{2^k}q^{2^{-k}+2^{k+s}}}
=\frac{v+uq^{2^s-2^{-k}}}{v+u^{2^k}q^{2^{k+s}-1}}=
({v+uq^{2^s-2^{-k}}})^{1-2^k},$$ 
which gives
\begin{equation}\label{eqL}
L_\frac{A}{B}\left(\frac{A}{B}x+x^{2^{-k}}\right)=0.
\end{equation}
Now $$ \frac{\Delta(x)}{B} = (\frac{A}{B}x + x^{2^{-k}}) +(\frac{C}{B}x^{2^s} + \frac{D}{B}x^{k+s}) =0 . $$
Applying this to Equation \ref{eqL} gives
$$L_\frac{A}{B}\left(\frac{\Delta(x)}{B}\right)=L_\frac{A}{B}\left(\frac{C}{B}x^{2^s}+\frac{D}{B}x^{2^{k+s}}\right)=0.$$
We compute this as
$$(B^{2^{-k}+{2^k}}C+D^{2^{-k}}A^{2^{k}+1})x^{2^s}+(B^{2^{-k}+{2^k}}D+B^{2^{-k}}AC^{2^k})x^{2^{k+s}}$$
$$+(B^{2^{-k}}AD^{2^k}+A^{2^k+1}C^{2^{-k}})x^{2^{-k+s}}=0.$$
We substitute in the values of $A, B, C, $ and $D$ and after
simplification we obtain the following
$$(vw+1)uq^{2^k+1+2^s}(vq^{2^{-k}}+uq^{2^s})(u^{2^k}q^{2^{k+s}+2^k}+u^{2^{-k}}q^{2^{-k+s}+1})x^{2^s}$$
$$+(vw+1)u^{2^k}q^{2^k+1+2^{k+s}}(vq^{2^{-k}}+uq^{2^s})(uq^{2^{k}+2^s}+u^{2^{-k}}q^{2^{-k+s}+2^{-k}})x^{2^{k+s}}$$
$$+(vw+1)u^{2^{-k}}q^{2^k+1+2^{-k+s}}(vq^{2^{-k}}+uq^{2^s})(u^{2^k}q^{2^{k+s}+2^{-k}}+uq^{2^s+1})x^{2^{-k+s}}=0.$$
As we chose $v$ and $w$ such that $v \neq w^{-1}$ and as $A \neq 0$
we can divide the equation by
$(vw+1)q^{2^k+1}(vq^{2^{-k}}+uq^{2^s})u^{2^{-k}+1}q^{2^{-k+s}+2^s+1}$
and take the expression to the $2^{-s}$ power to obtain
$$(1+a^{-2^{k-s}})x+(a^{2^{-s}}+a^{-2^{k-s}})x^{k}+(1+a^{2^{-s}})x^{2^{-k}}=0,  \ \ \ \ \  (1)$$
where $a=u^{2^{k}-1}q^{2^{-k}+2^{k+s}-2^{s}-1}.$ Now we consider
$L_\frac{C}{D}(\frac{\Delta(x)}{D})=0$. We know
$L_\frac{C}{D}(x^{2^s}+\frac{C}{D}x^{2^{k+s}})=0$, as
$$\frac{C}{D}=
\frac{wu^{2^{k}+1}q^{2^{k+s}+2^s}+uq^{2^s+1}}{wu^{2^k+1}q^{2^{k+s}+2^s}+u^{2^k}q^{2^{-k}+2^{k+s}}}
= ({w+u^{-1}q^{2^{-k}-2^s}})^{2^k-1}.$$ This implies
$L_\frac{C}{D}(\frac{A}{D}x+\frac{B}{D}x^{2^{-k}})=0$, which we
compute as
$$(C^{2^{-k}+2^k}A+C^{2^{-k}}DB^{2^k})x+(C^{2^{-k}}DA^{2^k}+D^{2^k+1}B^{2^{-k}})x^k$$
$$+(C^{2^{-k}+2^k}B+D^{2^k+1}A^{-k})x^{2^{-k}}=0.$$ A similar
computation to the one used above will yield
$$(1+a^{-2^{-k}})x+(1+a)x^{2^k}+(a+a^{-2^{-k}})x^{2^{-k}}=0. \ \ \ \ \  (2)$$
Now we combine equations $(1)$ and $(2)$ such that the terms in
$x^{2^{-k}}$ cancel. This will give
$$((1+a^{-2^{k-s}})(a+a^{-2^{-k}})+(1+a^{-2^{-k}})(1+a^{-s}))x+$$
$$((a^{2^{-s}}+a^{-2^{k-s}})(a+a^{-2^{-k}})+(1+a)(1+a^{-s}))x^{2^k}=0$$
which is the same as
$$((1+a^{-2^{k-s}})(a+a^{-2^{-k}})+(1+a^{-2^{-k}})(1+a^{2^{-s}}))(x+x^{2^k})=0.$$
If we show that
$(1+a^{-2^{k-s}})(a+a^{-2^{-k}})+(1+a^{-2^{-k}})(1+a^{2^{-s}}) \neq
0$ for all possible values of
 $a$ then we could conclude that $x \in GF(2^k)$.
To this end we consider the expression
$$(1+a^{-2^{k-s}})(a+a^{-2^{-k}})=(1+a^{-2^{-k}})(1+a^{2^{-s}}).$$
Rearranging we obtain
$$a=\frac{{(1+a^{-1})}^{2^{-k}}}{{(1+a^{-1})}^{2^{k-s}}}\frac{{(1+a)}^{2^{-s}}}{{(1+a)}^{2^k}}.$$
This implies $a$ is a $(2^{k+s}-1)$-th power which in turn implies
that it is a seventh power. As
$a=u^{2^{k}-1}q^{2^{-k}+2^{k+s}-2^{s}-1}=u^{2^{k}-1}q^{(2^{k+s}-1)(1-2^{-k})}$
we see that if $a$ is a seventh power then so is $u^{2^{k}-1}$ but
this is not possible as $k$ is not divisible by three and $u$ is
primitive. We can now state that all solutions to $\Delta(x)=0$ are
in $GF(2^k)$. Applying this to our original expression for
$\Delta(x)$ gives
$$(uq^{2^s+1}+u^{2^k}q^{2^{-k}+2^{k+s}})(x+x^{2^{s}})=0.$$
If $uq^{2^s+1}+u^{2^k}q^{2^{-k}+2^{k+s}}=0$ then $a=1$, but $1$ is a
seventh power, hence $(x+x^{2^{s}})=0$ which implies $x=0$ or $1$ as
$s$ is relatively prime to $3k$.

\section{Equivalence}

It remains to show that the new family of APN functions introduced
in this paper is indeed ``new".
We therefore need to demonstrate that these functions are not
CCZ equivalent to any known APN function.
Unfortunately no techniques currently exist for proving this by hand,
and we resort to a demonstration by computer for small values of $n$.
We attempt to show that the corresponding error-correcting codes
are inequivalent, which is necessary and sufficient as we said
in the introduction, and is proved in \cite{BBMMcG}.

One standard method of proving two codes to be inequivalent is
to show that they have a different weight distribution (if this is the case).
However, all the evidence shows that these codes all have the same weight distribution
as the code for the function $x^3$
(we have proved this for Family 5 in \cite{BBMMcG2}).
We will use other invariants.

Our quadrinomial Family 7

$$F(x)=u^{2^k}x^{2^{-k}+2^{k+s}}+ux^{2^{s}+1}+vx^{2^{-k}+1}+wu^{{2^k}+1}x^{2^{k+s}+2^s}$$

actually contains as a special case three of the  families
listed in the introduction, two of which are already known.
These are the binomial Family 1 when $v=w=0$, and
the trinomial Family 5 when $v=0, w\not= 0$.
Family 7 also contains Family 6 as a special case.
We claim that these four families are pairwise CCZ inequivalent.

 For smaller dimensions than 12, CCZ equivalence can be directly determined by
  testing equivalence of the asssociated codes with Magma.
For the case $n=6$ the polynomials introduced here take one of the following forms:

$$u x^3 +vu^5 x^{10} +v x^{17} +u^4 x^{24} $$
$$u x^3 +v x^{17} +u^4 x^{24} $$
$$ u x^3 +vu^5 x^{10} +u^4 x^{24}$$
$$u x^3 + u^4 x^{24}, $$

for some primitive element $u \in GF(2^{6})$ and $v \in GF(4)$.
In the first 3 cases, the polynomials are CCZ equivalent to
$$x^3+x^{10}+ u x^{24} ,$$  which appears in Dillon's list, and in the last instance the polynomial
is CCZ equivalent to $x^3$.
Therefore, $n=6$ is not a sufficiently large value of $n$ to distinguish our four families,
but does distinguish family 1 from families 5,6,7.

The next smallest possible value of $n$ to consider is $n=12$, so $k=4$.
Example functions (with $s=5$) from the four families are in the following table.

\bigskip

\begin{tabular}{|c|c|}
\hline
Function & Class   \\
\hline
                                                &                     \\
 $u^{16}x^{768}+ux^{33}+x^{257}+u^{290}x^{544}$ & Theorem \ref{thmain}\\
                                                & NEW (Family 7)  \\

\hline
                                 &      \\
$u^{16}x^{768}+ux^{33}+x^{257}$ & Theorem \ref{thmain} with $v\not=0$, $w=0$\\
                                 & NEW (Family 6)\\
\hline
                                   &           \\
$u^{16}x^{768}+ux^{33}+u^{290}x^{544}$ & Theorem \ref{thmain} \\
                                     &     with $v=0$, $w\not=0$ (Family 5) \\
\hline
                                     &           \\
$u^{16}x^{768}+ux^{33} $& Theorem \ref{thmain}\\
                                & $v=w=0$ (Family 1) \\
\hline
\end{tabular}
\bigskip

Magma has a built in test for code equivalence, which is sufficient for $n<12$.   This test involves
  performing a backtrack search using the action of the automorphism
  group of the code on the words of minimum weight.  However, for $n=12$  each
  of these codes has 1,397,760 words of miniumum weight and this is
  beyond the capability of the Leon package PERM for code equivalence
  that is used in Magma and other systems.

  John Cannon, Gabi Nebe and Allan Steel proved these codes to
  be inequivalent using a different approach. Firstly, the delta 2-rank
  of the four APN functions was determined.  The first three
  functions were found to have delta 2-rank 7900 while the fourth
  has delta 2-rank 7816.  Hence the fourth APN function is CCZ
  inequivalent to the first three.  All four functions were then
  shown to be pairwise CCZ inequivalent using a new invariant based
  on combinatorial properties of the words of minimum weight of
  the codes. All computations were done using Magma. We refer the
  reader to \cite{CNS} for details.

In conclusion, \cite{CNS} shows that our APN functions are new.

\bigskip

{\bf Acknowledgements}
We thank John Cannon, Gabriele Nebe, and Allan Steel 
for their work on APN functions and Magma.

\end{document}